\documentclass[conference,compsoc]{IEEEtran}
\usepackage[table]{xcolor}
\usepackage{booktabs}
\usepackage{tikz}
\usepackage[utf8]{inputenc}
\usepackage{graphicx}
\usepackage{caption}
\usepackage{subcaption}
\usepackage{scalerel}

\ifCLASSOPTIONcompsoc
  \usepackage[nocompress]{cite}
\else
  \usepackage{cite}
\fi
\ifCLASSINFOpdf

  \usepackage{pgfplots}
  \usepackage{CJK}
  \usepackage{pifont}
  \usepackage{lipsum}
  \usepackage{xcolor}
  \usepackage{multirow}
  \usepackage{tikz,pgfplots,pgfplotstable}
  
\else
\fi
\definecolor{bblue}{HTML}{4F81BD}
\definecolor{rred}{HTML}{C0504D}
\definecolor{ggreen}{HTML}{66CDAA}
\definecolor{ppurple}{HTML}{9F4C7C}
\definecolor{lightgray}{rgb}{0.93, 0.93, 0.93}

\begin{document}

%
\title{Simple But Not Secure: An Empirical Security Analysis of Two-factor Authentication Systems}



\author{
\IEEEauthorblockN{Zhi Wang}
\IEEEauthorblockA{DISSec, College of Cyber Science\\
Nankai University\\
Tianjin, China 300350\\
Email: zwang@nankai.edu.cn}
\and
\IEEEauthorblockN{Xin Yang}
\IEEEauthorblockA{DISSec, College of Cyber Science\\
Nankai University\\
Tianjin, China 300350\\
Email: yx@mail.nankai.edu.cn}
\and
\IEEEauthorblockN{Du Chen}
\IEEEauthorblockA{DISSec, College of Cyber Science\\
Nankai University\\
Tianjin, China 300350\\
Email: chendu@mail.nankai.edu.cn}
\and
\IEEEauthorblockN{Han Gao}
\IEEEauthorblockA{DISSec, College of Cyber Science\\
Nankai University\\
Tianjin, China 300350\\
Email: hangao@mail.nankai.edu.cn}
\and
\IEEEauthorblockN{Meiqi Tian}
\IEEEauthorblockA{Information Hub, Artificial Intelligence Thrust\\
The Hong Kong University of Science and Technology (Guangzhou) \\
Guangzhou, China 510000\\
Email: mtian837@connect.hkust-gz.edu.cn}
\and
\IEEEauthorblockN{Yan Jia}
\IEEEauthorblockA{DISSec, College of Cyber Science\\
Nankai University\\
Tianjin, China 300350\\
Email: jiay@nankai.edu.cn}
\and
\IEEEauthorblockN{Wanpeng Li\IEEEauthorrefmark{1}}
\IEEEauthorblockA{Department of Computing Science\\
University of Aberdeen\\
Email: wanpeng.li@abdn.ac.uk\\
\IEEEauthorrefmark{1}Corresponding author
}
}


%


\maketitle

\begin{abstract}

To protect users from data breaches and phishing attacks, service providers typically implement two-factor authentication (2FA) to add an extra layer of security against suspicious login attempts. However, since 2FA can sometimes hinder user experience by introducing additional steps, many websites aim to reduce inconvenience by minimizing the frequency of 2FA prompts. One approach to achieve this is by storing the user's ``Remember the Device'' preference in a cookie. As a result, users are only prompted for 2FA when this cookie expires or if they log in from a new device.

To understand and improve the security of 2FA systems in real-world settings, we propose SE2FA, a vulnerability evaluation framework designed to detect vulnerabilities in 2FA systems. This framework enables us to analyze the security of 407 2FA systems across popular websites from the Tranco Top 10,000 list. Our analysis and evaluation found three zero-day vulnerabilities on three service providers that could allow an attacker to access a victim's account without possessing the victim's second authentication factor, thereby bypassing 2FA protections entirely. A further investigation found that these vulnerabilities stem from design choices aimed at simplifying 2FA for users but that unintentionally reduce its security effectiveness. We have disclosed these findings to the affected websites and assisted them in mitigating the risks. Based on the insights from this research, we provide practical recommendations for countermeasures to strengthen 2FA security and address these newly identified threats.

\end{abstract}


%
\IEEEpeerreviewmaketitle

\section{Introduction}

With the rapid expansion of web services, implementing robust security measures to protect sensitive account data has become increasingly essential.
The Identity Theft Resource Center (ITRC) report reveals that data breaches have surged by 78\% in 2023, with a total of 3,205 incidents~\cite{ITRC}. 
Additionally, ``Identification and Authentication Failures'' remains a critical issue on the OWASP Top 10 list~\cite{owaspOWASPOWASP}, highlighting that account hijacking continues to be a significant threat to user security.

Two-factor authentication (2FA) is an enhanced security measure that helps protect accounts from unauthorized access by requiring an additional verification step beyond a password. 
By introducing a second authentication factor, 2FA mitigates risks associated with credential theft, as logging in requires more than just a password. This added layer of security makes unauthorized access significantly more challenging for attackers. 
For example, by the end of 2021, Google had automatically enabled two-step verification (2SV) for 150 million accounts, resulting in a reported 50\% decrease in account theft incidents~\cite{blogMakingSafer}.
While 2FA strengthens security by protecting organizations against account breaches, it also adds an extra verification step that can impact user experience. A key usability challenge with 2FA lies in task efficiency, such as the time needed to register or log in~\cite{katsini2016security}.

To enhance usability and reduce friction from the additional step, many online services use risk-based controls to selectively prompt 2FA. 
These mechanisms assess the security of each login environment, requiring 2FA only when a login attempt appears potentially risky.
For example, some websites offer a ``Remember the Device'' option during login. 
Once users successfully complete 2FA, the website will designate these devices as ``trusted'' for a set period, treating any future login attempts from them as risk-free.
This means users will not be asked to provide a second authentication factor when logging in from these ``trusted'' devices, thus improving 2FA’s usability by reducing repetitive prompts.

Websites rely on browser cookies to store information about ``trusted'' devices as part of their risk assessment processes in their 2FA systems.
These \textit{2FA Cookies} serve as indicators for future logins, allowing sites to bypass 2FA for recognized ``trusted'' devices.
While cookies themselves are not inherently insecure, improper configuration or flawed design can introduce vulnerabilities in these 2FA systems.
If 2FA cookies lack secure attribute settings, attackers could exploit various attack vectors — such as cross-site scripting (XSS) — to steal these cookies from victims~\cite{sivakorn2016cracked}. 
If an attacker gains access to a user’s 2FA cookies, they can impersonate the victim’s ``trusted'' device and bypass the second authentication factor required at login. 
Additionally, if the 2FA cookie is merely a flag bit or has a predictable value, an attacker could easily forge a valid 2FA cookie. This would allow them to bypass the 2FA protections set by the websites, effectively compromising the security intended by the systems.


Ensuring the security of 2FA is vital, especially as websites with large user bases increasingly support and promote 2FA activation. 
While previous studies have explored the use of browser fingerprints for enhancing authentication~\cite{acar2014web, ren2023secure, jiang2024two, reynolds2020empirical}, 
there has been limited focus on 2FA systems that use cookies to streamline the authentication process by reducing the frequency of 2FA prompts.

To address this gap, we propose \textbf{SE2FA}, an \textbf{S}ecurity \textbf{E}valuation framework for \textbf{2FA} systems, which allow us to conduct an in-depth empirical security analysis of real-world 2FA implementations, with a particular focus on websites that use cookies to streamline 2FA prompts and reduce authentication friction.
First, we examine 2FA support across popular websites from the Tranco Top 10,000 list, paying particular attention to the presence of ``Remember the Device'' functionality. 
To verify website risk control factors and capture 2FA cookies,
we developed a browser extension to assist in identifying these 2FA cookies on ``trusted'' devices.
Next, we evaluate the security of these 2FA cookies in a real-world environment by simulating an attacker who uses forged or stolen 2FA cookies to deceive the server into recognizing a device as ``trusted'', thereby bypassing the website's 2FA protections.
Finally, we evaluate the attribute configurations and design principles of 2FA cookies, identifying potential security weaknesses and their implications.

Our experiments reveal that most websites supporting 2FA and offering a ``Remember the Device'' feature use browser cookies as part of the user authentication process.
In practice, we found that the 2FA cookies on many sites can be exploited by attackers to bypass 2FA protections.
Additionally, we discovered that some websites fail to follow secure development practices when setting 2FA cookie attributes~\cite{bortz2011origin}, making them vulnerable to various attacks.
Lastly, we identified significant design flaws in the implementation of 2FA cookies on three websites, allowing attackers to easily forge or guess the 2FA cookies, thereby bypassing the 2FA systems with minimal effort.


In summary, we make the following contributions:

\begin{itemize}
    \item We propose a security evaluation framework for 2FA systems, \textbf{SE2FA}, which allows us to conduct an in-depth empirical security analysis of real-world 2FA implementations across the Tranco top 10,000 list, with a particular focus on websites that use cookies to implement the ``Remember the Device'' functionality within their 2FA system.
    \item Using SE2FA, we identified 377 popular websites from the Tranco top 10,000 list that support 2FA but were missing from the 2FA Directory dataset. Through the Directory’s contribution system, we submitted these findings to enhance the dataset, benefiting the broader community and supporting ongoing research on 2FA\@. 
    \item Our security analysis discovered three zero-day vulnerabilities in 93 websites relying solely on cookies to implement their ``Remember the Device'' function. These vulnerabilities allow attackers to forge or reuse 2FA cookies, effectively bypassing 2FA challenges and gaining access to a victim's account—without needing the victim's actual 2FA cookies.
    
    \item Our security analysis found that 52\% (93 out of 180) of the websites rely exclusively on cookies to implement the ``Remember the Device'' feature in their 2FA systems. This reliance exposes them to various vulnerabilities, including A1, A2, and A3 attacks, as detailed in Section \ref{sec:attacks}.
    \item We also provide recommendations for service providers implementing or planning to implement a ``Remember the Device'' feature within their 2FA systems. These guidelines aim to strengthen 2FA security and help prevent similar vulnerabilities in future deployments.
    
    
\end{itemize}

\textbf{Ethics Concerns.} Given the severity of our findings, we responsibly disclosed our results to the affected vendors.
We shared our methods and provided a demonstration video through the site's designated security contact.
It is important to note that all our experiments were conducted using test accounts and personal devices.
We did not interact with or affect any other users, nor did we attempt to attack real accounts.
After completing our tests, we logged out of all test accounts that allowed us to do so, in order to minimize any potential impact on the website's services.
To prevent any misuse of the identified vulnerabilities, we have anonymized the names of the sites we believe to be affected in this paper.


\section{Problem and Approach Overview}

In this section, we first provide an overview of two-factor authentication (2FA) systems. Next, we describe specific security issues in 2FA systems that implement the ``Remember the Device'' feature. Finally, we discuss the challenges involved in conducting an empirical security analysis of 2FA implementations and outline our approach to addressing these challenges.

\subsection{Two Factor Authentication}

Knowledge-based authentication systems, which rely on a username and password combination, are susceptible to phishing attacks, brute force attacks, and data breaches~\cite{huang2011using, mirante2013understanding, dastane2020effect}. 
To address these risks, service providers are increasingly shifting from knowledge-based, single-factor authentication (SFA) ~\cite{bruun2014usability} toward multi-factor authentication (MFA), which provides enhanced security. Among MFA options, two-factor authentication (2FA) as the most widely adopted form, has been implemented by various service providers to strengthen account security~\cite{10.1007/978-981-16-4244-9_23}.

Since 2011, the number of service providers offering 2FA has grown dramatically, rising from just three providers in 2011 to 910 within the top 10,000 providers in our study (see Table \ref{tab:2fa-dataset}).
Recognizing the importance of 2FA, Google has declared that a second layer of authentication is essential for account security. In 2021, Google automatically enabled two-step verification for 150 million accounts, resulting in at least a 50\% reduction in account compromise risk~\cite{ozaki2024operation}.

Two-factor authentication requires an additional verification factor beyond the standard username and password. One-time passwords (OTP) are the most commonly used factor in 2FA systems, providing an additional layer of security against unauthorized access, even if a password is compromised~\cite{eldefrawy2011otp}.
OTPs are generated either periodically or in response to each authentication attempt, using a seed and other factors assigned to the user during the registration process\cite{hsieh2011design, sun2015trustotp}. 
In practice, the service providers complete the second authentication step by verifying the OTP supplied by the user. A variety of methods may be employed by users to receive or generate OTPs, including the use of authenticator applications~\cite{berrios2023factorizing}, Short Message Service (SMS), phone calls, hardware tokens, and email.

\subsection{Observation and Attacks}

\subsubsection{Risk Control Measures for 2FA System}
\label{subsub:RiskControl}

To prevent unauthorized access, service providers often employ various risk control measures designed to detect potentially fraudulent login attempts~\cite{wiefling2019really}. However, to improve user experience and enhance usability, many websites offer a ``Remember the Device'' option, allowing users to designate specific devices as ``trusted''~\cite{patat2020please}. This feature enables users to bypass the 2FA prompts when logging in from the same device within a valid period. In the following section, we outline the risk control measures commonly used by service providers to implement the ``Remember the Device'' feature in their 2FA systems.


\textbf{Cookie-Based Risk Control Measure.} 
A widely used approach for risk control in 2FA systems is the use of browser cookies ~\cite{patat2020please}, a technology natively supported by web browsers. Cookies are stored on the user's device and are typically employed to remember user information and interactions with websites. In 2FA systems, particularly those that implement the ``Remember the Device'' feature, cookies are also employed to identify ``trusted'' devices~\cite{jacomme2021extensive}.
For example, some websites use a cookie named \texttt{2fa\_devices} to mark a device as ``trusted''. When this cookie is included in the login request, the server recognizes the device as trusted, allowing the user to bypass the 2FA prompts during subsequent logins.

\textbf{Fingerprint-Based Risk Control Measure.} 
Browser fingerprinting is a technique used to uniquely identify users by collecting a variety of characteristics related to their browser and device~\cite{7546540, vastel2018fp, 10.1007/978-3-319-99136-8_26}. This measure allows for identification and tracking without relying on cookies. Due to its high accuracy and representativeness, browser fingerprinting is integrated into the risk control in 2FA systems to enhance usability~\cite{lin2022phish}.
During the login process, the website generates a fingerprint of the user's browser and sends it to the server. The server compares this fingerprint with those of previously ``trusted'' devices. If a match is found, the device is deemed ``trusted'', enabling the user to bypass the 2FA prompt during login.

\textbf{Other Risk Control Measures.} 
During our security analysis of 2FA system implementations, we identified one website
employ \texttt{localStorage}~\cite{mozillaWindowLocalStorage} to retain information for identifying ``trusted'' devices within the user's browser (see Section \ref{subsec:security-risks-rq2}). 
Additionally, some websites incorporate the user's IP address as part of the risk control mechanisms for their 2FA systems, requiring 2FA challenges even when users log in from the same device but with a different IP address.
Some websites combine multiple risk control measures to enhance the ``Remember the Device'' feature and improve the usability of their 2FA systems. For example, \texttt{planfix.com} uses both browser cookies and browser fingerprints to verify trusted devices. In this case, users must log in using the same browser with the ``Remember the Device'' option selected previously to avoid 2FA prompts during subsequent logins. 


\subsubsection{Observation}

It was observed that while the original intention behind the ``Remember the Device'' feature in 2FA is to enhance user experience and reduce the friction caused by frequent 2FA prompts, this improvement in usability can inadvertently compromise the security of 2FA systems if proper attention is not given during its implementation.
Vulnerabilities in this feature could allow an attacker to bypass the 2FA protections, rendering the system ineffective and undermining its intended security benefits. 
Lin et al.~\cite{LinISP22} showed that attackers could steal user browser fingerprints through phishing attacks, 
deceiving service providers into believing that login attempts are coming from a ``trusted'' device. 
This allows attackers to bypass 2FA and gain unauthorized access to the victim's account. 
However, there is currently a lack of comprehensive security analysis of the 2FA ``Remember the Device'' implementations that rely on cookies, as well as the potential risks and vulnerabilities associated with this approach.

To address this gap, we reviewed 10,000 websites from the top 10,000 in Tranco's list~\footnote{https://tranco-list.eu/, assessed 04-07-2024} and identified 910 online services that support 2FA.
Among these, we found 180 websites that implemented a ``Remember the Device'' feature during the login process. We developed a comprehensive experimental approach to evaluate websites that use browser cookies to support this functionality.
Our research reveals that more than half of these sites rely exclusively on 2FA cookies to store information about ``trusted'' devices.
Furthermore, we highlight the security vulnerabilities and potential attack vectors associated with using browser cookies as the sole factor in risk-based authentication.

\subsubsection{Attacks}

\label{sec:attacks}

To evaluate the security of 2FA systems with the  ``Remember the Device'' feature that relies on cookies to identify ``trusted'' devices, we propose the following adversary model. This model explores how an attacker could bypass cookie-based 2FA mechanisms. We assume that the attacker has already gained access to the victim's first authentication factor — typically a username and password — through phishing attacks or by exploiting a data breach that exposes web service account credentials~\cite{huang2011using, dastane2020effect}.
In this threat model, we assume that the attacker is able to acquire or forge the 2FA cookies set by the targeted website. This could be achieved by exploiting vulnerabilities in the cookie handling process or by taking advantage of logic flaws in the design of the 2FA systems. With both the victim's username and password, along with the stolen or forged 2FA cookies, the attacker can bypass the 2FA, thus gaining direct access to the victim's account. Once inside, the attacker can take control of the account, potentially establishing persistent access for continued exploitation~\cite{dao2021alternative}.

In this paper, we consider four types of attackers who could potentially obtain the victim's 2FA cookies.
Each of these attackers can be identified in real-world scenarios.
We now describe four specific attacks that these attackers could carry out.

\vspace{0.15cm}
\noindent\textbf{(A1)} Man-in-the-Middle (MITM) Attack: A \textbf{network attacker} could steal a user's cookies if the \texttt{SECURE} flag is either absent or incorrectly set to false.
The \texttt{SECURE} flag is intended to ensure that cookies are only transmitted over encrypted HTTPS connections, protecting them from interception over unencrypted HTTP channels\cite{sivakorn2016s}.
If the \texttt{SECURE} flag of the user's 2FA cookies is absent or misconfigured, a network attacker could intercept the victim's 2FA cookies during transmission using an MITM attack\cite{saltzman2009active}.

\vspace{0.15cm}
\noindent\textbf{(A2)} Cross-Site Scripting (XSS) Attack: A \textbf{web attacker} can exploit XSS vulnerabilities in websites to steal cookies, especially if the \texttt{HTTPOnly} flag is either absent or incorrectly set to false~\cite{rodriguez2018cookie}. The \texttt{HTTPOnly} flag is designed to prevent client-side scripts from accessing cookie data, adding an additional layer of security. However, if this flag is misconfigured, an attacker could use techniques such as script injection to steal the victim's 2FA cookies.

\vspace{0.15cm}
\noindent\textbf{(A3)} Man-in-the-browser(MITB) Attack: A \textbf{MITB attacker}~\cite{owaspManinthebrowserAttack} can gain control of a victim's browser by using Trojans or malicious browser extensions~\cite{pantelaios2020you, nayak2024experimental}.
This allows the attacker to intercept and modify data sent between the browser and the server, including cookies.
Malicious browser extensions can read cookies directly from the victim's browser and transmit them to the attacker~\cite{tyler2024towards, saini2016colluding}.
This enables the attacker to retrieve all cookies from the victim's browser, including the 2FA cookies, regardless of the cookie's security flags.

\vspace{0.15cm}
\noindent\textbf{(A4)} Logic Flaw Attack: Our analysis reveals that some websites have design flaws in their 2FA cookies implementation, allowing these cookies to be reused or easily predicted.
In this scenario, the attacker acts as an \textbf{observer}. 
By monitoring the 2FA cookies set by the website, the attacker can either use his own 2FA cookies or forge valid 2FA cookies to disable the 2FA protection on the victim's account and gain unauthorized access to the victim's account.

\subsection{Challenges and Our Approach}

\noindent Conducting a large-scale empirical security analysis of 2FA implementations presents the following three challenges. 

\vspace{0.15cm}
\noindent\textbf{C1}: What is the most efficient way to determine if a given website supports 2FA?
The current dataset for tracking 2FA implementations across popular websites is the 2FA Directory~\cite{2faDirectory}, an open-source project that aggregates a list of websites supporting 2FA\@. 
However, this dataset relies on community contributions, and our evaluation (see Section \ref{sec:2fa-support}) has revealed that a significant number of websites supporting 2FA are missing in this dataset. 
Additionally, manually reviewing 10,000 websites would be excessively time-consuming. Therefore, it is crucial to develop an automatic method that can quickly identify whether a website supports 2FA, improving the efficiency of subsequent testing and reducing the need to manually review websites that do not offer 2FA\@.

\vspace{0.15cm}
\noindent\textbf{C2}: What methodology can be employed to determine whether the ``Remember the Device'' feature is implemented using browser cookies, and to retrieve the corresponding 2FA cookies?
Only developers of a website have full knowledge of the technical specifics of their 2FA implementation. For researchers, it is often unclear whether a website’s ``Remember the Device'' feature relies on cookies, browser fingerprints, or other risk control measures to implement its 2FA system. Therefore, it is essential to develop a comprehensive and accurate approach to test whether a website uses cookies to identify ``trusted'' devices, facilitating accurate security analysis of these 2FA systems.


\vspace{0.15cm}
\noindent\textbf{C3}: How can the authenticity of obtained or forged 2FA cookies be verified in a real-world operational context, and can they be leveraged to bypass the victim's 2FA challenge?
Some websites may use a combination of techniques to identify ``trusted'' devices in their 2FA systems.  To evaluate whether obtained or forged 2FA cookies can bypass a victim's 2FA challenge, it is crucial to simulate realistic attack scenarios. This includes evaluating cookies' ability to circumvent 2FA protections while considering other security mechanisms like IP address verification and browser fingerprinting, which may influence the success or failure of the attack. 



To solve these challenges, we designed the \textbf{SE2FA} vulnerability evaluation framework to analyze potential vulnerabilities within 2FA systems effectively.
To address the first challenge, we identify websites supporting 2FA and retrieve their 2FA documentation by utilizing a meta-search engine along with advanced search syntax.
And then, we focused on websites implementing a ``Remember the Device'' feature, applying our evaluation method to detect cookie-based implementations and retrieve 2FA cookies. 
At last, to validate the security impact of 2FA cookies, we simulated a realistic attack environment in which an attacker attempts to use 2FA cookies to bypass the victim's 2FA protections. This allowed us to assess whether the website might have underlying security weaknesses or vulnerabilities.


\section{Design}
\label{sec:design}

\begin{figure*}[ht]
    \centering
    \includegraphics[width=\textwidth]{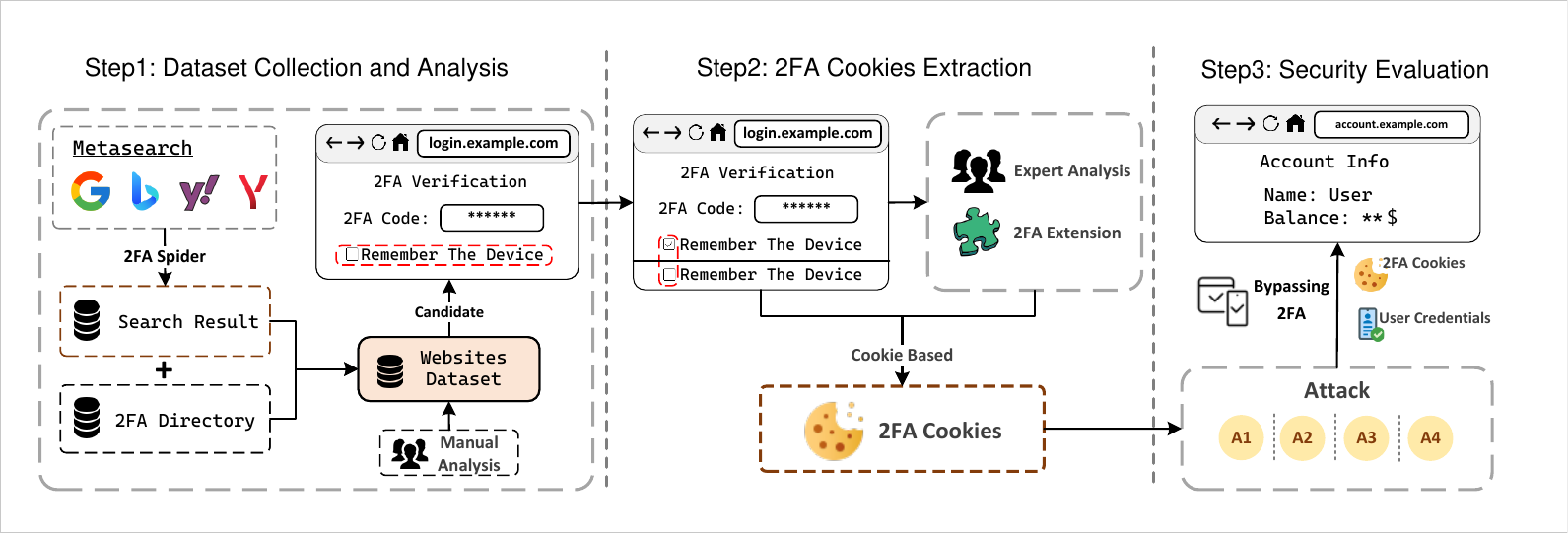}
    \caption{\textbf{System Architecture of SE2FA}. SE2FA initially gathers 2FA information from websites using meta-search engines, analyzing whether each site supports a ``Remember the Device'' feature. It then extracts the website’s 2FA cookies via 2FA extension and manual review. Finally, SE2FA evaluates the security of the website's 2FA implementation within a simulated attack environment.}
    \label{fig:2FA-All}
\end{figure*}

As shown in Figure \ref{fig:2FA-All}, our approach begins by using a meta-search engine to gather a list of websites that support 2FA\@. This step is crucial because many websites that support 2FA are not included in the 2FA Directory dataset. The initial results from the meta-search engine are then combined with the top 10,000 websites listed in the 2FA Directory that support 2FA to form our dataset.  
Next, each website in this dataset is manually examined to determine if it allows registration, supports 2FA, and offers a ``Remember the Device'' feature. 
For websites meeting these criteria, we developed a browser extension, called 2FA Extension, to collect 2FA cookies from websites that use cookie-based 2FA implementations.
Finally, the collected 2FA cookies are verified and analyzed in a real-world environment to assess whether the websites may have potential security vulnerabilities.

\subsection{Dataset Collection and 2FA Coverage Analysis}
\label{subsec:dataset-collection}

SE2FA first uses a meta-search engine to collect websites from the top 10,000 list of Tranco that support 2FA, as demonstrated in Figure \ref{fig:2FA-All}. While the 2FA Directory provides data on websites supporting 2FA within the top 10,000 websites, it relies on community contribution. As a result, many websites that do support 2FA are missing in this dataset. 
To build a comprehensive dataset that contains all websites supporting 2FA within the top 10,000 websites, we developed a crawler, named \textbf{2FA Spider}\footnote{https://anonymous.4open.science/r/SE2FA-541F/}, to automate the process of collecting websites potentially supporting 2FA\@. 
2FA Spider queries the open-source search engine SearXNG~\cite{githubGitHubSearxngsearxng} with the search phrase ``2FA OR MFA website'' alongside the domain name of each website. 
It then collects results from four search engines (Google, Bing, Yandex, and Yahoo) to identify the most relevant candidates based on keyword matching.
To address the issue of false positives in search results, 2FA Spider includes a filtering mechanism. It sets a threshold to exclude results that are not specifically related to 2FA for the searched website, ensuring more accurate data collection.

Using 2FA Spider, we identified 1,371 websites from the top 10,000 that potentially support 2FA. Due to false positivities from meta searches,  we manually reviewed and excluded websites without login interfaces or those lacking 2FA support, resulting in a refined set of 798 websites with 2FA\@. 
We also retained websites with restricted registration (e.g., educational sites) to assess 2FA adoption among popular websites, as detailed in Section \ref{sec:2fa-support}.

Our results matched 421 of the 533 2FA-enabled websites in the 2FA Directory, yielding an accuracy of 79\% (see Table \ref{tab:2fa-dataset}). This demonstrates that our tool can effectively identify websites with 2FA support, significantly reducing the need for manual verification.

We then added 112 websites, which 2FA Spider missed but were present in the 2FA Directory, to our dataset, resulting in a final set of 910 2FA-supported websites. After further verification, we identified an additional 377 websites supporting 2FA that were not listed in the 2FA Directory. Excluding educational, government, and other restricted websites, we ultimately contributed 103 new 2FA-supported websites to the 2FA Directory through its contribution system, aiming to benefit the broader community and support ongoing 2FA research. 

\begin{table}[t]
    \centering
    \resizebox{0.5\textwidth}{!}{ 
        \rowcolors{1}{white}{lightgray}
        \begin{tabular}{cccccc}
            \toprule
            \textbf{2FA Directory($D_1$)} & \textbf{2FA Spider($D_2$)} & \textbf{$D_1-D_2$} & \textbf{$D_2- D_1$} & \textbf{$D_1\cap D_2$} &\textbf{Total}\\
            \midrule
            533 & 798 & 112 & \textbf{377} & 421 & \textbf{910} \\
            \midrule
        \end{tabular}
    }
    \caption{\textbf{Comparison of 2FA Directory and Our 2FA Spider.} $D_1-D_2$  represents websites in the 2FA Directory not found by our 2FA Spider, while $D_1\cap D_2$ shows the matches identified by 2FA Spider. $D_2- D_1$ refers to websites identified by our 2FA Spider that are not included in the 2FA Directory. } 
     \label{tab:2fa-dataset}
\end{table}

Therefore, we manually analyzed these 910 websites to assess whether they offer a registration interface, support 2FA, and include the ``Remember the Device'' function. We categorized these websites into five groups:

\begin{itemize}
    \item G1: Websites with registration and 2FA support, but without the ``Remember the Device'' feature.
    \item G2: Websites with registration, 2FA support, and the ``Remember the Device'' feature.
    \item G3: Websites that require third-party accounts, e.g., ``outlook.com'' requires a Microsoft account for login.
    \item G4: Websites that cannot be registered (e.g., educational websites).
    \item G5: Websites with registration and 2FA support, but cannot be enabled for some reason (e.g., only subscription accounts can enable 2FA).
\end{itemize}

The categorized data is shown in Table \ref{tab:2fa-grouped}. We identified 180 websites that fall into the G2 category. For each of these websites, we created two accounts to conduct testing and verification, ensuring that 2FA was enabled on both accounts. All websites in the G3 category actually rely on the 2FA systems in G2, and the detailed data is shown in Appendix \ref{sec:appendix-third-party} Table \ref{tab:third-party}.



\begin{table}[!h]
    \centering
    \resizebox{0.5\textwidth}{!}{
        \setlength{\tabcolsep}{8pt}
        \renewcommand{\arraystretch}{1.3}
        \begin{tabular}{c|c|c|c}
            \hline
            \textbf{Group} & \textbf{Description} & \multicolumn{2}{c}{\textbf{Number}} \\
            \hline
            \rowcolor{lightgray}
            \textbf{G1} & 2FA without ``Remember the Device'' & \multicolumn{2}{c}{227}\\
            \hline
            \multirow{2}{*}{\textbf{G2}} & 2FA with ``Remember the Device'' (\textbf{Cookie-Only Measure}) & \textbf{93} & \multirow{2}{*}{180} \\
                                & 2FA with ``Remember the Device'' (Other Measures) & 87 & \\
            \hline
            \rowcolor{lightgray}
            \textbf{G3} & Require third-party accounts & \multicolumn{2}{c}{62} \\
            \hline
            \textbf{G4} &  Cannot be registered & \multicolumn{2}{c}{430} \\
            \hline
            \rowcolor{lightgray}
           \textbf{G5} & 2FA cannot be enabled & \multicolumn{2}{c}{11} \\
            \hline
        \end{tabular}
    }
    \caption{Number of Websites with 2FA Support and Associated Risk Control Measures.}
    \label{tab:2fa-grouped}
\end{table}


\subsection{2FA Cookie Extraction}

In this step, we use a combination of automated tools and manual review to extract 2FA cookies from target websites (see step 2 in Figure \ref{fig:2FA-All}). 
Websites often use a variety of cookies for different purposes, and there are currently no specialized tools designed for 2FA cookie analysis, making fully automated analysis a significant challenge.
Furthermore, the diverse ways in which websites implement and present their 2FA systems further complicate the task.
To address these challenges, we adopt a hybrid approach that combines cookie extraction tools with human expert review, ensuring a comprehensive and accurate assessment of 2FA cookies.

To achieve this, we developed a custom browser extension, called \textbf{2FA Extension}~\footnote{https://anonymous.4open.science/r/SE2FA-541F/}, designed to meet the specific requirements of our analysis. 2FA Extension could allow us  
to capture cookies from the current browser environment and the visited website at a given moment, creating what we refer to as a ``cookie snapshot''. 
The extension also enables us to compare two different cookie snapshots, highlighting any changes, additions, or deletions, including changes to cookie attributes.
Additionally, it provides an intuitive interface for selectively enabling or disabling individual cookies, allowing us to focus on those most relevant to our analysis.
To maintain a clean testing environment between experiments, the extension includes functionality to clear browser data (including cookies, webpage data, history, etc.).
Furthermore, by using the \texttt{webRequest} permission~\cite{chromeWebRequest}, 2FA-Extension provides a network packet capture function, allowing us to analyze HTTP headers (particularly the \texttt{Cookie} and \texttt{Set-Cookie} headers) in both HTTP requests and responses~\cite{ietf2616Hypertext} during our experiments.

However, the presence of 2FA cookies does not necessarily imply that the website relies exclusively on cookies to verify ``trusted'' devices.
To determine if the website employs alternative risk control measures to secure its 2FA implementation, we move on to the next stage of verification.


\subsection{Security Evaluation}
\label{sec:security-evaluation}


In this stage (see step 3 in Figure \ref{fig:2FA-All}), we first assess whether the website relies solely on 2FA Cookies to identify ``trusted'' devices or if it uses additional authentication methods (see Section \ref{subsub:RiskControl}).
We also check whether the website notifies the user if a new login attempt has been detected.

To determine whether a website relies solely on 2FA cookies to identify ``trusted'' devices, we first import the cookies obtained from the target website in step 2 to a new device. This device should have a different IP address and browser, ensuring a distinct browser fingerprint. We then initiate the login process using the victim's username and password. If no 2FA prompt appears on the new device, we can conclude that the website relies entirely on cookies to implement its ``Remember the Device'' feature. Such websites could be vulnerable to attacks (such as A1, A2, A3, and A4), as discussed in Section \ref{sec:attacks}.




To assess if the victim receives a notification for login attempts from a new device, we monitor the victim’s registered email and phone number after attempting the 2FA bypass. If no notifications (emails or messages) are received, we can conclude that the victim may remain unaware of the attacker’s actions, leaving the account potentially exposed to undetected 2FA bypass attacks.


Additionally, we evaluated the security issues associated with 2FA cookies. 
If the attributes of these cookies are not configured properly, they could expose the 2FA system to attacks such as \textbf{A1} and \textbf{A2}. 
Moreover, we believe that relying solely on 2FA cookies to implement the ``Remember the Device'' function is inherently insecure, as it leaves the system vulnerable to the \textbf{A3} attack as described in Section \ref{sec:attacks}. 

Finally, we identified serious design flaws in the 2FA cookie implementation on three websites. These flaws arise from using fixed or predictable values for the 2FA cookies, potentially allowing attackers to reuse or forge valid 2FA cookies. This vulnerability could enable attackers to bypass the 2FA prompt on a victim's account without the need for the victim's 2FA cookies (see \textbf{A4} in Section \ref{sec:attacks}).



\section{Evaluation}
\label{sec:evalution}
To assess the security of 2FA systems on websites and the effectiveness of our attacks on real-world online services, we address the following research questions:

\textbf{RQ1 (Adoption Study)}: How many popular websites currently support 2FA, what methods do they use for implementation, and how many of these websites offer a ``Remember the Device'' feature?

\textbf{RQ2 (Vulnerability Detection)}: What vulnerabilities are present in 2FA implementations on websites that rely on cookies to identify ``trusted'' devices, and are these sites vulnerable to the attacks we present?


\subsection{Experiment Setup}



In our evaluation, we established two distinct login environments: one for the victim account and one for the attacker. The configurations of these environments are as follows:

\begin{itemize}
    \item \textbf{Victim's Setup:} 
    \begin{itemize}
        \item A Windows 10 laptop using the Chrome browser (version 123.0.6312.87).
        \item A smartphone with commonly used authenticator apps installed for 2FA\@.
    \end{itemize}
    \item \textbf{Attacker's Setup:}
    \begin{itemize}
        \item A Windows 11 laptop running the Firefox browser (version 124.0.2).
        \item A smartphone with commonly used authenticator apps installed for 2FA\@.
    \end{itemize}
    \item \textbf{Network and Fingerprinting Distinctions:} The two laptops were assigned unique IP addresses and placed in different geographical locations. Their browser fingerprints were verified as distinct using FingerprintJS~\cite{githubGitHubFingerprintjsfingerprintjs}.
\end{itemize}



\subsection{Testing Procedure}
\label{sec:remember-the-device}

In this section, we outline the methodology for conducting our large-scale empirical security analysis of real-world 2FA systems.

\vspace{0.15cm}
\noindent\textbf{Account Registration.} 
For the 910 websites (see Table \ref{tab:2fa-grouped}) implementing a 2FA system, we created two distinct user accounts (one for each environment) for the websites we were able to register with and enabled 2FA on both accounts using the respective smartphones. The registration and 2FA setup processes were carefully executed to ensure compatibility with each website’s normal functionality. 
It is important to note that, due to the nature of interacting with critical 2FA challenges on the target websites, this process required the involvement of human expertise for completion and could not be automated.


\vspace{0.15cm}
\noindent
\textbf{Identifying ``Remember the Device''.} 
After enabling 2FA, we then checked whether the website provided a ``Remember the Device'' or similar option. This feature typically appears in one of three places:

\begin{itemize}
    \item During Login Verification: Some websites offer a ``Remember the Device'' or ``Trust this Browser'' option after completing the second factor of authentication.
    \item In 2FA Settings: Some websites allow users to mark a device as ``trusted'' when configuring 2FA settings.
    \item Alternative Mechanisms: Features like ``Remember Me'', which keep users logged in without requiring 2FA challenges, are also considered.
\end{itemize}

It is important to note that due to the variation in how 2FA interfaces are implemented, designing an automated method to detect this feature is highly challenging. As a result, we relied on human experts to manually identify its presence. In our analysis, we have identified 180 websites that provide a ``Remember the Device'' (see G2 in Table \ref{tab:2fa-grouped}).

\vspace{0.15cm}
\noindent
\textbf{Identifying Risk Control Measures.} 
When a website offers a ``Remember the Device'' feature, it is not always clear to us what risk control measures are employed by the website to implement this functionality. To detect the risk control mechanisms in use, we conduct the following tests:

\begin{enumerate}
    \item First, we log into the target website using the victim's account with the ``Remember the Device'' option selected (as shown in step 2 of Figure~\ref{fig:2FA-All}).
    \item Next, we log out and clear all browser data, including cookies, to eliminate any potential interference from previous sessions. 
    \item We then log back into the victim's account. If a new 2FA prompt appears, we conclude that the website uses cookies as part of its risk control for the 2FA system.
\end{enumerate}

At this stage, it remains unclear whether additional risk control measures are used in conjunction with the cookie-based one. To investigate further, we conduct the following additional test:

\begin{enumerate}
    \item First, we log into the target website using the victim's account with the ``Remember the Device'' option selected (see step 2 in Figure~\ref{fig:2FA-All}).
    \item We then log out from the victim's account on the victim's browser.
    \item Next, we export the cookies from the victim's browser and import them onto the attacker's machine using our 2FA Extension.
    \item Finally, we attempt to log into the victim's account on the attacker's machine. If a new 2FA prompt appears, we conclude that the website employs additional risk control measures, beyond just cookies, as part of its 2FA system.
\end{enumerate}

Since the login process involves solving challenges from the 2FA system, we rely on human expertise to complete the tests described above with the assistance of our 2FA Extension. Upon completing these tests, we observed that 93 websites rely solely on browser cookies to implement their ``Remember the Device'' feature. 



\vspace{0.15cm}
\noindent
\textbf{Identifying Cookies Used for ``Remember the Device''.} For websites that rely solely on cookie-based risk control measure for the ``Remember the Device'' feature, we utilize the cookie comparison functionality of our 2FA Extension to identify differences between two sets of cookies: one from the ``Remember the Device'' state and another from the non-``Remember the Device'' state.

In most cases, we observe that the ``Remember the Device'' state includes additional cookies, such as a \texttt{trustedDevice} cookie. 
However, some websites may rely on multiple cookies to identify the trusted device and might not use meaningful names for these cookies, making it difficult to detect the actual cookies being used. In such cases, we leverage the enable/disable cookie functionality of our 2FA Extension to identify the exact cookies required by the website. This process involves manually enabling each cookie one by one to determine which cookies are crucial for identifying the ``trusted'' device.



\vspace{0.15cm}
\noindent
\textbf{Evaluating the Effectiveness of Cookie-Stealing Attacks.} 
Up until now, we have identified the exact cookies used by the 93 websites to support their 2FA systems (see Table \ref{tab:2fa-grouped}). To verify that the identified cookies can indeed be used to bypass the 2FA challenge on the target website, and to minimize the possibility of false positives, we perform the following test:
\begin{enumerate}
    \item First, we log into the target website using the victim's account with the ``Remember the Device'' option selected (as shown in Step 2 of Figure~\ref{fig:2FA-All}).
    \item Next, we export only the 2FA-related cookies (excluding all other cookies) from the victim's browser and import them onto the attacker's machine using our 2FA Extension. 
    \item Finally, we attempt to log into the victim's account on the attacker's machine. If no 2FA prompt appears, we confirm that the exported cookies from Step 2 are indeed the 2FA cookies and can be used to bypass the 2FA challenge.
\end{enumerate}

Once we confirm that the 2FA cookies can bypass the target website's 2FA challenge, we proceed to assess the security measures placed on these cookies using our 2FA Extension. Specifically, we examine security flags such as \texttt{Secure} and \texttt{HttpOnly}, and also check the expiration times of the 2FA cookies. These details help us determine which types of attacks (e.g., A1, A2, and A3 in Section \ref{sec:attacks}) the target website's 2FA system might be vulnerable to.



\vspace{0.15cm}
\noindent
\textbf{Design Flaw Evaluation.} 
When a website relies solely on cookies to implement its ``Remember the Device'' feature, a design flaw can arise if insufficient attention is given to how the cookie values are generated. For example, if the 2FA cookie is simply a flag or is predictable, minimal effort is required for an attacker to forge or reuse a cookie and bypass the target website's 2FA challenge. To identify such design flaws, we conduct the following test:

\begin{enumerate}
    \item First, we log into the target website using the two accounts we registered, selecting the ``Remember the Device'' option on both the victim's machine and the attacker's machine, respectively.
    \item Next, we export the 2FA cookies from both machines.
    \item We repeat Steps 1 and 2 to collect four sets of 2FA cookies: two sets from the attacker's account and two sets from the victim's account.
    \item We then compare the four sets of 2FA cookies. If the values of the 2FA cookies are identical across all four sets, we immediately conclude that a design flaw exists in the target website's 2FA system. If the values differ, we investigate whether the cookies are fixed or predictable between login attempts and across accounts.
    \item For websites with fixed or predictable cookies, we test the vulnerability by forging the 2FA cookies on the attacker's machine and using them to log into the victim's account.
\end{enumerate}

Design flaws like this allow attackers to bypass the 2FA challenge without requiring access to the victim's two-factor authenticator, making the website vulnerable to account takeover attacks and unauthorized access. 



\vspace{0.15cm}
\noindent
\textbf{Detecting Notification of Suspicious Login Attempts.} 
During our evaluation, we also observed how the target websites handled the detection of new login attempts. Specifically, we monitored whether the websites issued alerts when an attacker used the victim's 2FA cookies to bypass the 2FA challenge.
We recorded the websites (see Table \ref{tab:2fa-notifications}) that sent email notifications or other alerts in response to these login attempts. 




\section{Results}
\label{sec:result}

Using SE2FA, as described in Section \ref{sec:design} and \ref{sec:evalution}, we evaluated the security of 2FA systems across websites in the Tranco top 10,000 list. Our study examined a total of 910 websites with 2FA support, with special emphasis on those offering a ``Remember the Device'' feature. For these sites, we conducted an in-depth security evaluation to assess potential vulnerabilities. In the following sections, we present the findings from our security analysis.


\subsection{Overall Status(RQ1)}

\subsubsection{2FA Support}
\label{sec:2fa-support}

\begin{figure}[h!]
    \centering
    \begin{subfigure}{0.45\textwidth}
        \includegraphics[width=\textwidth]{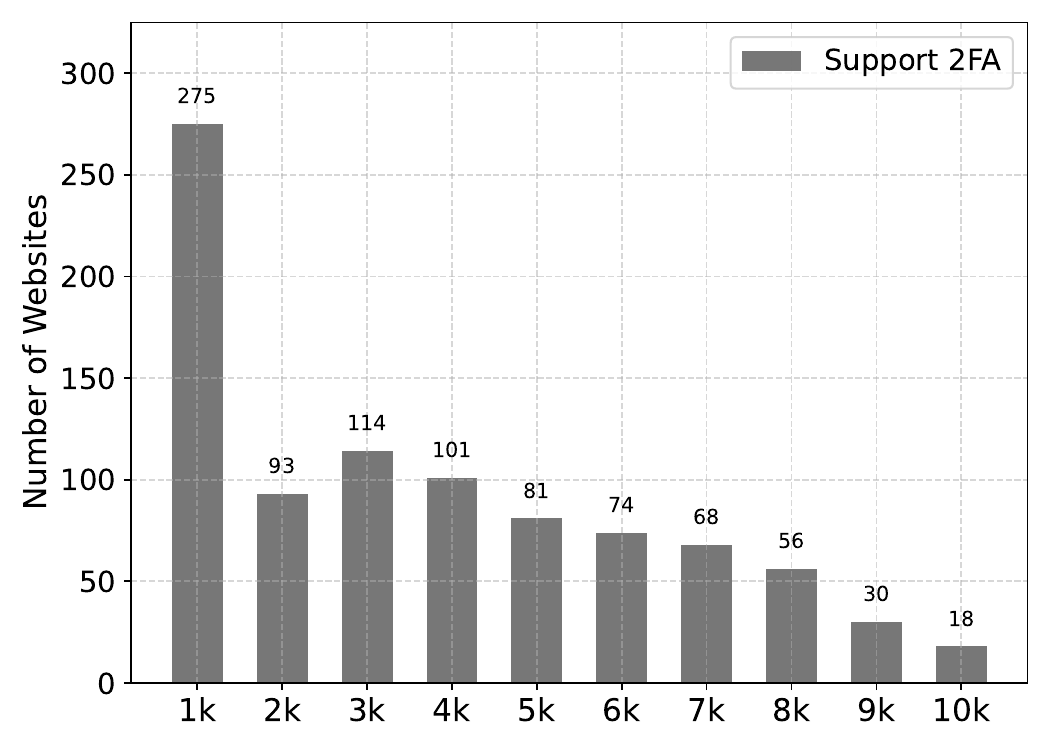}
        \caption{2FA Support By Ranking.}
        \label{fig:website_distribution}
    \end{subfigure}
    \hfill
    \begin{subfigure}{0.51\textwidth}
        \includegraphics[width=\textwidth]{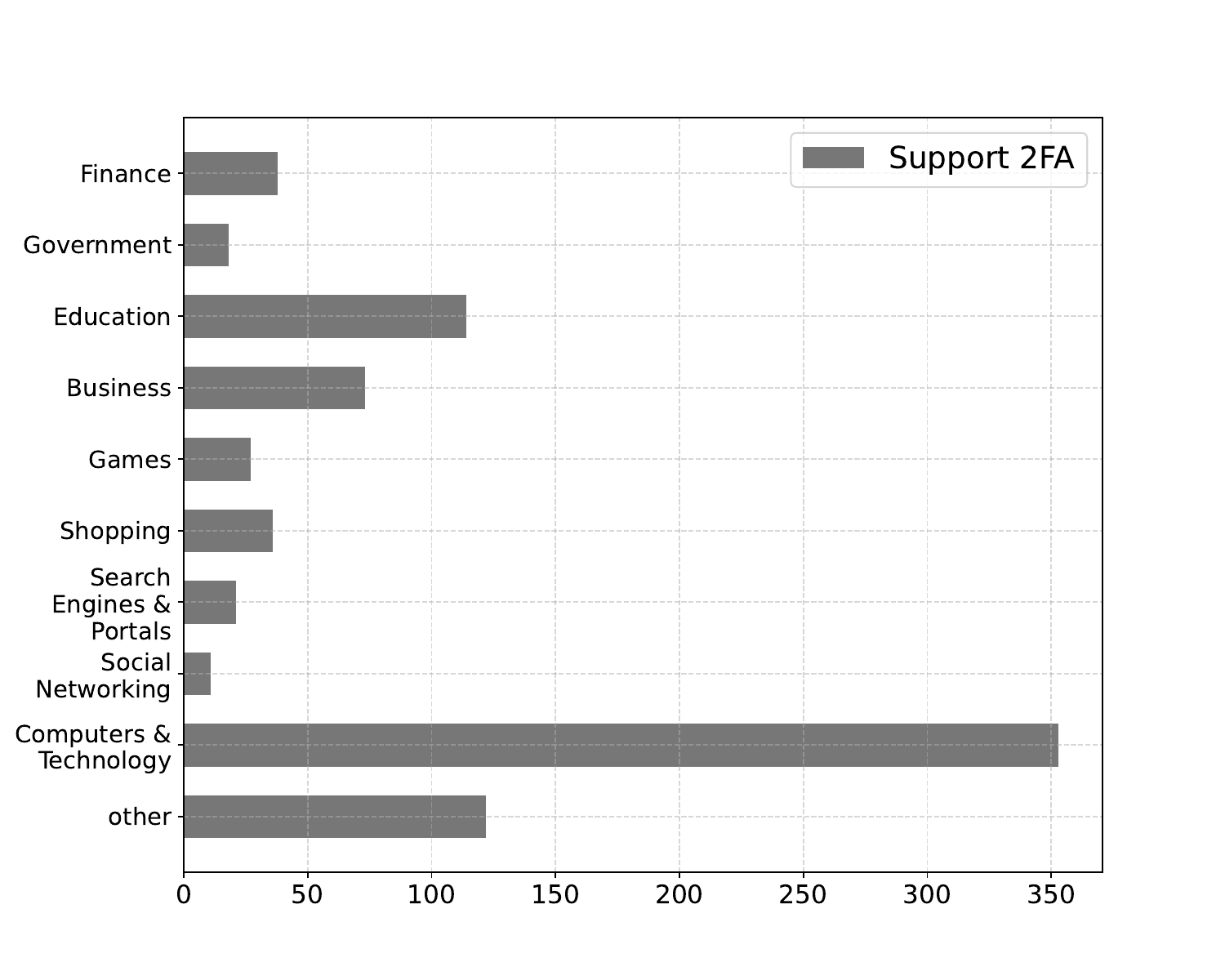}
        \caption{2FA Support by Website Category.}
        \label{fig:website_categories}
    \end{subfigure}
    \caption{2FA Adoption on Top 10,000 Websites.}
    \label{fig:2fa_support}
\end{figure}

Figure \ref{fig:website_distribution} shows the distribution of 2FA support across every 1,000 websites in the Tranco top 10,000,  categorized by their ranking. 
Our findings indicate that higher-ranked websites are more likely to employ 2FA to protect user accounts.
Additionally, our review also includes websites where we were unable to register, such as those belonging to educational institutions (see G3 in Table \ref{tab:2fa-grouped}). For these sites, we used our 2FA Spider to crawl search results and retrieve their 2FA documentation, confirming whether they offer 2FA support.



It is important to note that our investigation focused on determining whether popular online providers support 2FA. 
As a result, we excluded websites like \texttt{google.com.au} and retained the main site, \texttt{google.com}, since both point to the same underlying platform. Including both would not provide additional value for our analysis of 2FA cookies and website security.

In addition, we used the Cyren URL Category Checker~\footnote{https://data443.com/cyren-url-category-check-gate/} to categorize the tested websites and analyze 2FA support within each category.
For clarity, we grouped categories with fewer than 10 websites into an ``Other'' category, as shown in Figure \ref{fig:website_categories}. Our analysis reveals 
that websites in categories such as Computers \& Technology, Education, Business, Shopping, and Finance were more likely to support 2FA\@.
This trend is likely driven by the sensitive nature of the services they provide — such as financial transactions, asset storage, and online purchases — which require an additional layer of security to protect user accounts.


\subsubsection{2FA Verification Methods Used}

In our analysis, we identified nine common 2FA verification methods used by popular websites: SMS, phone call, app, hardware token, email, passkey, authenticator app, biometrics, and recovery code. 
Notably, we distinguish between authenticator apps and specific apps. 
The latter refers to apps related to a specific website (e.g., \texttt{so.com} requires verification via the 360 Mobile Guard app) that do not specifically generate a time-based one-time password (OTOP).
Authenticator apps, by contrast, are dedicated applications designed to generate OTOPs. Among the 910 websites we analyzed, the most common 2FA verification method is authenticator apps, used by 650 websites, as shown in Table \ref{tab:2fa-methods}.

Our findings also reveal that most websites offer only a single 2FA verification method. 
However, a few websites provide up to five different verification options.
Furthermore, we observed that websites offering multiple 2FA verification methods generally favor authenticator apps, followed by SMS and email, as these verification methods are the most convenient for users, as shown in Table \ref{tab:2fa-methods}. 
Additionally, a small number of sites also support alternative verification methods such as biometrics, hardware tokens, passkeys, and recovery codes.

\begin{table}[t]
    \centering
    \setlength{\tabcolsep}{8pt} 
    \renewcommand{\arraystretch}{1.3} 
    \rowcolors{1}{white}{lightgray} 
    \begin{tabular}{c|c|ccccc} 
        \hline
        \multicolumn{2}{c|}{\textbf{2FA Verification Methods}} & \multicolumn{5}{c}{\textbf{Number of Methods Supported}} \\ 
        \hline
        \textbf{Method} & \textbf{Total} & 1 & 2 & 3 & 4 & 5 \\
        \hline
        SMS                & 330  & 46  & 126 & 99 & 52 & 7 \\
        Phone Call         & 116  & 4   & 19  & 40  & 46 & 7 \\
        Specific App       & 118  & 62  & 19  & 18  & 14 & 5 \\
        Hardware Token     & 77   & 2   & 17  & 32  & 21 & 5 \\
        Email              & 197  & 58  & 65  & 43  & 26 & 5 \\
        Passkey            & 19   & 0   & 6   & 6   & 4  & 3 \\
        Authenticator App  & 650  & 356 & 141 & 101 & 45 & 7 \\
        Biometrics         & 9    & 3   & 3   & 0   & 2  & 1 \\
        Recovery Code      & 16   & 2   & 6   & 6   & 2  & 0 \\
        \hline
    \end{tabular}
    \caption{The 2FA Verification Methods Used in Popular Websites.}
    \label{tab:2fa-methods}
\end{table}



\subsection{Security Risks Associated with Cookies(RQ2)}
\label{subsec:security-risks-rq2}


In Section \ref{sec:remember-the-device}, we analyzed a total of 407 websites and identified 180 websites that support 2FA and offer the ``Remember the Device'' feature.
During the review, we excluded websites in G4 and G5 because we were unable to register or enable 2FA on these sites. Additionally, we excluded websites in G3, as they all rely on the 2FA systems in G2 to authenticate users.

In total, we identified 93 websites that only use cookie-based methods to identify ``trusted'' devices. This means that 52\% (93 out of 180)  of the websites offering a ``Remember the Device'' option depend solely on 2FA cookies for device identification. On these websites, we also executed the attack described in Section \ref{sec:security-evaluation} to eliminate false positives. 

\begin{figure}
    \centering
    \includegraphics[width=8cm]{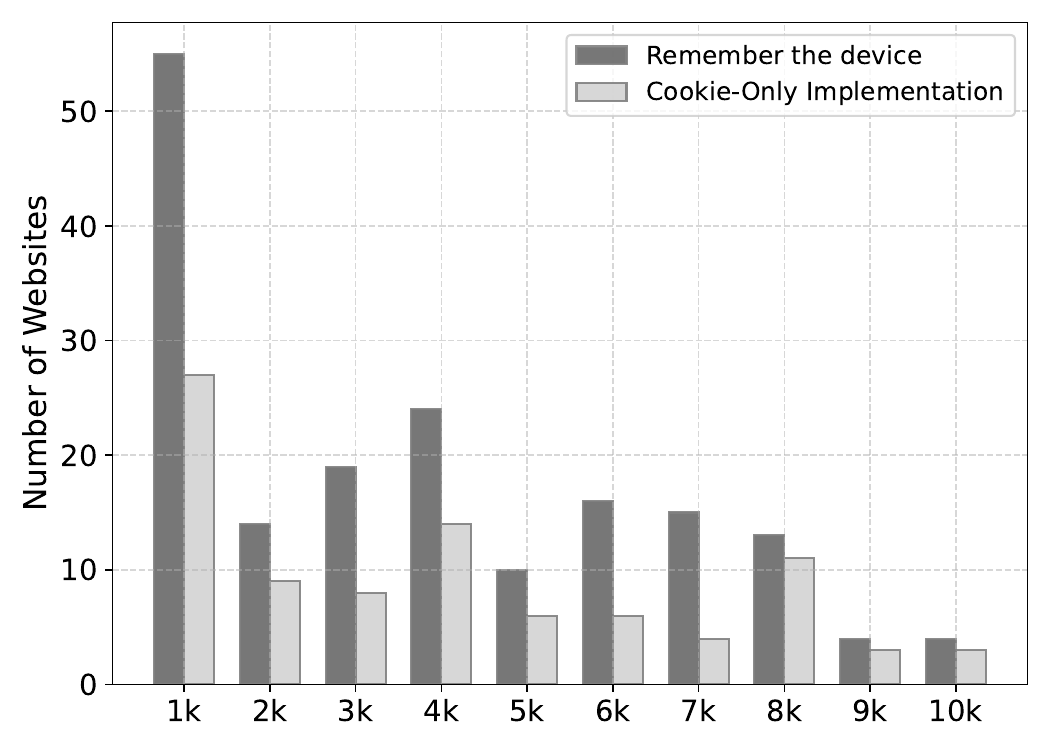}
    \caption{
Websites Supporting the ``Remember the Device'' Feature and Those Using Cookie-Only Implementations.}
    \label{fig:remember_cookie}
\end{figure}

In addition, we plotted the number of popular websites that offer the ``Remember the Device'' feature, as well as those that rely solely on 2FA cookies to identify ``trusted'' devices, as shown in Figure \ref{fig:remember_cookie}. 
The data reveals that higher-ranked websites, which are more popular, tend to use a combination of risk control measures (see Section \ref{subsub:RiskControl}) to identify ``trusted'' devices. 
In contrast, lower-ranked websites are more like to rely solely on cookies to implement the ``Remember the Device'' feature.

For the remaining websites, we observed that some use browser fingerprints, while others rely on a combination of cookies, fingerprints, and IP address detection to manage the risk control measures of their 2FA systems. 
Additionally, we found one website that stored ``trusted'' device information in the web page’s \texttt{localStorage}.
Similar to cookies, \texttt{localStorage} is a web storage mechanism that permanently stores data in key-value pairs within the user's browser until explicitly deleted~\cite{mozillaWindowLocalStorage}.
Since \texttt{localStorage} can also be accessed through XSS attacks, we consider it a special variant of 2FA cookies.

\subsubsection{Security Issues in 2FA Cookie Management}

After collecting the 2FA cookies from websites that rely solely on 2FA cookies to identify ``trusted'' devices, we analyzed the attribute settings of these cookies. 
Due to page limit, we present the cookies attributes and their associated attacks for the top 500 websites that only use cookies to protect their ``Remember the Device'' feature in Table \ref{tab2:top-500}. Data for the remaining websites can be found in Appendix Table \ref{tab:top10000}.

For websites that use multiple cookies to identify a trusted device, we focus on the most secure cookie, as an attacker would need to obtain all 2FA cookies to successfully bypass the 2FA challenge and access the victim’s account.
Given the seriousness of our findings, especially considering that accounts on these services are highly valuable and often targeted, we have chosen to anonymize the websites in our presentation. 
This includes anonymizing both the website names and the cookie names to protect their identities. 

\begin{table*}[t]
    \centering

    \begin{tabular}{c|c|cccc|c|c}
        \toprule
        \multirow{2}{*}{\textbf{No.}} & \multirow{2}{*}{\textbf{Website}} & \multicolumn{4}{c|}{\textbf{Cookie Attribute}} & \multirow{2}{*}{\textbf{Design Flaws (\textbf{A4}})} & \multirow{2}{*}{\textbf{Attack Type}} \\
        && Amount & HTTPOnly & Secure & Expiries (days) & & \\
        \midrule

        \rowcolor{lightgray}
        1 & Social Networking - 1 & 1 & \textcolor{black}{\ding{51}} & \textcolor{black}{\ding{51}} & 3 & - & \textbf{A3} \\
        2 & Computers \& Technology - 1 & 1 & \textcolor{black}{\ding{51}} & \textcolor{black}{\ding{51}} & 400 & - & \textbf{A3} \\
        \rowcolor{lightgray}
        3 & Professional Networking - 1 & 2 & \textcolor{black}{\ding{51}} & \textcolor{black}{\ding{51}} & 17 & - & \textbf{A3} \\
        4 & Shopping - 1 & 1 & \textcolor{black}{\ding{51}} & \textcolor{black}{\ding{51}} & 365 & - & \textbf{A3} \\
        \rowcolor{lightgray}
        5 & Personal StorageComputers \& Technology - 1 & 1 & \textcolor{black}{\ding{51}} & \textcolor{black}{\ding{51}} & 90 & - & \textbf{A3} \\
        6 & Computers \& Technology - 2 & 1 & \textcolor{black}{\ding{51}} & \textcolor{black}{\ding{51}} & 100 & - & \textbf{A3} \\
        \rowcolor{lightgray}
        7 & Computers \& Technology - 3 & 1 & \textcolor{black}{\ding{51}} & \textcolor{black}{\ding{51}} & 30 & - & \textbf{A3} \\
        8 & Games - 1 & 1 & \textcolor{black}{\ding{51}} & \textcolor{black}{\ding{51}} & 400 & - & \textbf{A3} \\
        \rowcolor{lightgray}
        9 & Games - 2 & 1 & \textcolor{black}{\ding{51}} & \textcolor{black}{\ding{51}} & 90 & - & \textbf{A3} \\
        10 & Search Engines \& Portals - 1 & 1 & \textcolor{red}{\ding{55}} & \textcolor{red}{\ding{55}} & 365 & - & \textbf{A1}, \textbf{A2}, \textbf{A3} \\
        \rowcolor{lightgray}
        11 & BusinessComputers \& Technology - 1 & 1 & \textcolor{black}{\ding{51}} & \textcolor{black}{\ding{51}} & 365 & - & \textbf{A3} \\
        12 & Personal Storage - 1 & 1 & \textcolor{black}{\ding{51}} & \textcolor{black}{\ding{51}} & 400 & - & \textbf{A3} \\
        \rowcolor{lightgray}
        13 & Computers \& Technology - 4 & 1 & \textcolor{black}{\ding{51}} & \textcolor{black}{\ding{51}} & 30 & - & \textbf{A3} \\
        14 & Business - 1 & 1 & \textcolor{black}{\ding{51}} & \textcolor{black}{\ding{51}} & 14 & - & \textbf{A3} \\
        \rowcolor{lightgray}
        15 & GamesForums \& Newsgroups - 1 & 1 & \textcolor{black}{\ding{51}} & \textcolor{black}{\ding{51}} & 30 & \ding{108} & \textbf{A3}, \textbf{A4} \\
        16 & Business - 2 & 1 & \textcolor{black}{\ding{51}} & \textcolor{black}{\ding{51}} & 180 & - & \textbf{A3} \\
        \rowcolor{lightgray}
        17 & Shopping - 2 & 1 & \textcolor{red}{\ding{55}} & \textcolor{black}{\ding{51}} & 365 & - & \textbf{A2}, \textbf{A3} \\
        18 & Games - 3 & 1 & \textcolor{black}{\ding{51}} & \textcolor{black}{\ding{51}} & 400 & - & \textbf{A3} \\
        \rowcolor{lightgray}
        19 & Business - 3 & 1 & \textcolor{black}{\ding{51}} & \textcolor{black}{\ding{51}} & 30 & - & \textbf{A3} \\
        20 & Computers \& Technology - 5 & 1 & \textcolor{red}{\ding{55}} & \textcolor{black}{\ding{51}} & 30 & - & \textbf{A2}, \textbf{A3} \\
        \rowcolor{lightgray}
        21 & Computers \& Technology - 6 & 1 & \textcolor{black}{\ding{51}} & \textcolor{black}{\ding{51}} & 90 & - & \textbf{A3} \\
        22 & Search Engines \& Portals - 2 & 1 & \textcolor{black}{\ding{51}} & \textcolor{black}{\ding{51}} & 30 & - & \textbf{A3} \\
        \rowcolor{lightgray}
        23 & Computers \& Technology - 7 & 1 & \textcolor{red}{\ding{55}} & \textcolor{black}{\ding{51}} & 30 & - & \textbf{A2}, \textbf{A3} \\
        \bottomrule
    \end{tabular}
    \caption{Top 500 Websites Vulnerable  to \textbf{A1, A2, A3} and \textbf{A4} Attacks. \ding{51}:2FA Cookie attribute set to \texttt{true}, \textcolor{red}{\ding{55}}: 2FA Cookie attribute set to \texttt{false}, \ding{108}: Design flaw - cross account reuse.} 
    \label{tab2:top-500}
\end{table*}

We found that the \texttt{Secure} attribute of 2FA cookies was not set on eight websites, making these cookies vulnerable to the \textbf{A1} attack we proposed.
As a result, MITM attackers can intercept and obtain 2FA cookies through MITM attacks.
Additionally, the \texttt{HttpOnly} attribute of 2FA cookies was not set on 11 websites, leaving these websites vulnerable to our \textbf{A2} attack, which allows web attackers to steal cookies through XSS attacks.
Furthermore, we believe that all websites implementing the ``Remember the Device'' feature based solely on 2FA cookies are inherently insecure.
This is because they are vulnerable to the \textbf{A3} attack we proposed, where attackers can infect users' browsers by distributing malicious browser extensions.
These malicious extensions can access all of the user's browser cookies, including 2FA cookies, bypassing the security attributes of the cookies.

We also evaluated the expiration time of the 2FA cookies we collected. Our findings show that only nine websites (10\%) set the expiration time of their 2FA cookies to 7 days or less. Less than half of the websites (48\%) set this attribute to less than 30 days, with the most common setting being 30 days (30\%). 14 websites even set the expiration time to 400 days.
A long cookie expiration time significantly increases the attack window. Once an attacker successfully steals the 2FA cookies, they can bypass the user's 2FA challenge for an extended period, potentially leading to long-term access to the victim’s account before the cookies expire. This extended window of vulnerability poses a serious security risk.

\subsection{Design Flaws(RQ2)}

In our security analysis, we identified three websites with critical design flaws in their 2FA implementations and two websites that are not properly implementing their 2FA systems.
Notably, on one of these websites, 2FA cookies could be used across multiple accounts when ``Remember the Device'' option is selected.
For example, on ``GamesForums \& Newsgroups - 1'', an attacker could bypass the 2FA challenge on a victim’s account using their own 2FA cookies, without needing the victim's cookies.
We also found that the ``Remember the Device'' feature on two websites was poorly designed, making the 2FA cookies predictable or easily forgeable.
For instance, one website set the \texttt{otp\_verified\_at} cookie to \texttt{timestamp} after the ``Remember the Device'' option was selected. An attacker could easily forge such a cookie and use it to gain unauthorized access to the victim's account. 
Additionally, we discovered that one website used Base64 instead of proper encryption to encode its 2FA cookies. 
Upon decoding these cookies, we found sensitive information related to the user, including the user's IP address, login date, and OTP input. 
Moreover, we identified two websites that claimed to support 2FA; however, despite enabling the 2FA feature in their system, no 2FA challenge appeared during any login attempt. We believe these websites are not properly implementing their 2FA systems.


In total, we identified three zero-day vulnerabilities on three websites that would allow an attacker to bypass the 2FA challenge on the victim account.  We responsibly disclosed these vulnerabilities to the affected websites. 
While one website initially claimed this behavior was a ``feature'', we verified in subsequent testing that they had implemented a fix following our report.

\section{Discussion}

Two-factor authentication has received increasing attention from service providers, with approximately 28\% of the Tranco Top 1,000 websites now supporting it (see Figure \ref{fig:website_distribution}). Although 2FA enhances protection against phishing and credential-stuffing attacks, it does come with usability trade-offs. Each login requires the user to complete a 2FA challenge, and frequent prompts can lead to frustration, sometimes causing users to disable 2FA altogether. To strike a balance between security and user convenience,  around 44\% of these websites (180 out of 407) have implemented a ``Remember the Device'' feature in their 2FA systems.

\subsection{Lessons learned}
Most of the vulnerabilities described in Section \ref{sec:result} are caused by how service providers handle 2FA cookies and the design choices made for managing these cookies. We next consider in greater detail how and why the various classes of vulnerability that we have identified have arisen.

\subsubsection{Improperly Managed 2FA Cookies} When a website relies only on cookies to manage risk controls for its 2FA system, careful handling of these cookies is crucial for preventing security vulnerabilities. However, our security analysis found that 15\% of these websites (14 out of 93) did not follow the best practices to set the \texttt{Secure} and \texttt{HttpOnly} attributes.
Additionally, more than 80\% (74 out of 93)of these websites set their cookie expiration period to 30 days or longer, with 30 websites even setting the expiration time to one year. 

Setting a long expiration period or omitting an expiration time entirely for 2FA cookies further widens the window of vulnerability.  If an attacker gains access to a user’s 2FA cookies via an XSS or network sniffing attack, they can bypass the 2FA challenge and potentially maintain unauthorized access to the account until the cookies eventually expire. For example, the ``Search Engines \& Portals - 1'' website (see Table \ref{tab2:top-500}) set the expiration time for their 2FA cookies to 365 days and also left both \texttt{Secure} and \texttt{HttpOnly} flags disabled.

\subsubsection{Simplicity Beats Security}
\label{subsubsec:simplicity}

Implementing a feature like ``Remember the Device'' within a 2FA system can require significant development effort. However, developers are often under pressure to meet tight project deadlines, which can lead to shortcuts in implementation. To save time, they may focus on delivering the functionality quickly while overlooking critical security protection needed for this feature.
As a result, our analysis uncovered three zero-day vulnerabilities across the evaluated websites, highlighting how prioritizing simplicity over security can introduce serious security risks to 2FA systems. These vulnerabilities expose user accounts to potential unauthorized access, putting sensitive information at risk.

\subsubsection{Put all Eggs in One Basket}
\label{subsubsec:eggs}
As discussed in Section \ref{subsub:RiskControl}, various measures could be used to manage risk controls in 2FA systems. The ``Remember the Device'' feature should be implemented without sacrificing the security of the underlying 2FA system. However, our security analysis revealed that 52\% (93 out of 180) of these websites choose to only use cookies to manage the ``Remember the Device'' feature in their 2FA systems. This exclusive reliance on cookies introduces significant security vulnerabilities. Without additional protection, attackers can exploit these cookies through attacks like XSS, network sniffing and man in the browser, allowing unauthorized access to accounts by bypassing 2FA protections.

\subsection{Mitigation}


Our security analysis has revealed severe vulnerabilities in 2FA systems that offer a ``Remember the Device'' feature. 
There is a significant danger that these vulnerabilities will be replicated in future systems.
Below, we make a number of recommendations, directed at service providers, designed to address the vulnerabilities we have identified. There are two reasons for making these recommendations, namely both to try to address the problems that exist in current systems and to help ensure that future systems are built in a more robust way.

\subsubsection{Best Practices for Securing 2FA Cookie Implementation} 
Implementing robust cookie security practices is essential to protect 2FA cookies from unauthorized access. We have the following recommendations:


\begin{itemize}
    \item \textbf{Set} \texttt{Secure} \textbf{and} \texttt{HttpOnly} \textbf{Attributes}: Enable both the \texttt{Secure} and \texttt{HttpOnly} attributes for 2FA cookies. The \texttt{Secure} attribute prevents cookies from being transmitted over insecure HTTP connections, mitigating the risk of network sniffing (A1). The \texttt{HttpOnly} attribute restricts access to cookies via JavaScript, helping to protect against XSS vulnerabilities (A2).
    \item \textbf{Set a Short Expiration Time}:  Limiting the lifespan of 2FA cookies reduces the risk of stolen cookies being reused by attackers. For enhanced security, configure 2FA cookies to expire within a few days or less, rather than allowing them to persist for extended periods. 
\end{itemize}

\subsubsection{Combining Multiple Risk Control Measures} Our security analysis reveals that relying solely on cookies for risk management in 2FA systems is insufficient and inherently insecure. 
We recommend that service providers combine the following risk control measures when implementing the ``Remember the Device'' feature:

\label{sec:security-methods}

\begin{itemize}
    \item \textbf{Browser Fingerprinting}: By capturing non-invasive data, such as browser settings and device characteristics, websites can generate a unique device identifier for the user. When a login attempt is detected from an unrecognized device, this identifier allows the service provider to assess the risk and, if necessary, trigger a 2FA challenge to verify the user's identity.
    \item \textbf{IP Address and Geolocation}: Including a user's IP address and location in the risk assessment can add another layer of verification. If the location or IP is inconsistent with previous logins, the website can trigger an additional security check.
    \item \textbf{Behavioral Analytics}: Analyzing user behavior, such as login frequency and device switching, helps detect anomalies and triggers extra authentication if necessary.
\end{itemize}

Although implementing multiple risk control measures may increase development complexity, it significantly improves user account security, reducing the likelihood of unauthorized access even if user credentials are compromised.

\subsubsection{User Notifications for New Login Attempts} 
\label{subsubsec:user-notifications}

In our security analysis, 45 websites promptly notify users of suspicious login behavior by sending an email alert for new or unusual logins, as shown in Table \ref{tab:2fa-notifications}.
This security feature effectively alerts users in real time, allowing them to respond promptly to potentially unauthorized access attempts. 
Additionally, some websites automatically trigger a password reset in response to failed authentication attempts or other suspicious activity, safeguarding accounts from unauthorized access. Table~\ref{tab:2fa-notifications} provides an overview of the types of email notifications sent by websites when a user's account is accessed from a new device in our simulated attack environment. 

\begin{table}[!h]
    \centering
    \resizebox{0.5\textwidth}{!}{
        \setlength{\tabcolsep}{8pt}
        \renewcommand{\arraystretch}{1.3}
        \rowcolors{1}{white}{lightgray}
        \begin{tabular}{>{\centering\arraybackslash}m{1.2cm} >{\centering\arraybackslash}m{6cm} >{\centering\arraybackslash}m{1.2cm}}
            \toprule
            \textbf{Notification Type} & \textbf{Description} & \textbf{Number} \\
            \midrule
            N1 & New device login notification only & 24 \\
            N2 & New device login time and location notification & 12 \\
            N3 & Abnormal IP login notification & 5 \\
            N4 & Suspicious login verification & 2 \\
            N5 & Unauthorized login attempt notification and automatic password reset & 1 \\
            N6 & Notification that the 2FA Code is incorrect but the account password is correct & 1 \\
            \midrule
        \end{tabular}
    }
    \caption{Types of Email Notifications Sent When a New Login Attempt is Detected.}
    \label{tab:2fa-notifications}
\end{table}



\section{Related Work}
To the best of our knowledge, this paper presents the first comprehensive security analysis of real-world two-factor authentication (2FA) systems that use browser cookies as part of their risk management strategies. We now review the related work in this section.


\subsection{Cookie Security.}

Drakonakis et al.~\cite{DrakonakisIP20} developed an automated auditing framework to detect authentication and authorization flaws in web apps related to cookie handling, revealing significant vulnerabilities in session hijacking and cookie hijacking across 25K domains. Despite HTTPS adoption, many domains lack proper HSTS deployment, making them susceptible to eavesdropping and hijacking attacks. Calzavara et al.~\cite{CalzavaraUTSS21} analyzed client-enforced security policies across 15,000 popular sites, revealing widespread inconsistencies in cookie security attributes, CSP, and HSTS configurations, leading to vulnerabilities like XSS and cookie theft. They found that the current Origin Policy proposal is insufficient to address these issues and propose a new Site Policy framework to ensure explicit security measures, supported by a publicly available prototype and toolchain. Khodayari et al.~\cite{KhodayariP22} conducted the first comprehensive security evaluation of the SameSite cookie policy, analyzing its usage trends, the impact of the new default policy, and associated threats. They concluded that while SameSite cookies can significantly reduce XS attack surfaces, their effectiveness depends on proper implementation and developer awareness. Gavazzi et al.~\cite{GavazziWKLKDL23} studied 208 popular sites and found that a minority support MFA and RBA, leaving many users vulnerable to account hijacking. However, using SSO providers with MFA and RBA significantly improves security, though it introduces privacy concerns due to third-party tracking.

\subsection{2FA Security.}

\textbf{Usability and User Experience in 2FA.} To enhance usability, Karapanos et al.~\cite{KarapanosMSC15} proposed a transparent 2FA mechanism that verifies user proximity through ambient noise recordings from the phone and computer. Reynolds et al.~\cite{ReynoldsSRDRS18} conducted studies showing that novice users face significant challenges setting up YubiKeys for 2FA, but find them highly usable for daily authentication once properly configured. Golla et al.~\cite{GollaHLPR21} conducted large-scale controlled experiments on Facebook to examine how messaging and UX design patterns can improve 2FA adoption, identifying best practices for encouraging protective behavior and suggesting future directions for promoting digital security through effective security prompts. Lyastani et al.~\cite{Lyastani0B23} examined the consistency of 2FA user experiences on top-ranked websites, finding minimal design consistency and problematic user experience elements suggesting the need for general UX guidelines for 2FA implementation. Smith et al.~\cite{SmithYDRS23} proposes a  secondary authentication factor 
(SAF) manager to assist users with the lifecycle of secondary authentication factors in 2FA. Through user studies, the SAF manager was shown to improve the setup, removal, and replacement of SAFs, preventing fatal errors and receiving positive user feedback. Daffalla and Bohuk et al.~\cite{DaffallaBDBR23} are the first to evaluate the effectiveness of user-facing security interfaces on online services, finding can hide or spoof access details in device lists and session logs.  Klemmer et al.~\cite{KlemmerGSASF23} analyzed developer advice on authentication, identifying significant challenges and suggesting improvements to enhance the usability and security of web authentication.

\textbf{Security Vulnerabilities and Risk Analysis in 2FA.} Shrestha et al.~\cite{ShresthaSSS16} examined the security of second-factor login based on ambient audio, highlighting a vulnerability where an attacker can exploit predictable or known sounds from the phone (e.g., ringer, notifications) to bypass the system, as these sounds dominate the phone's recordings over ambient noise. Sun et al.~\cite{sun2015trustotp} presented a one-time password solution that leverages ARM TrustZone technology et al.~\cite{Alves2004TrustZoneI} to combine the flexibility of software tokens with the security of hardware tokens, ensuring OTP protection even if the mobile OS is compromised. Reaves et al.~\cite{ReavesSTBTB16}conducted the first longitudinal security study of the modern text messaging ecosystem, uncovering how companies implement SMS-based security services and how defenses like phone-verified accounts are circumvented. Thomas et al.~\cite{ThomasLZBRIMCEM17} conducted a longitudinal study of the underground credential theft ecosystem, uncovering millions of potential victims and highlighting the effectiveness of enhanced authentication mechanisms in mitigating account hijacking. Lin et al.~\cite{LinISP22} explored the security risks of using browser fingerprints for authentication, demonstrating that attackers can replicate users' fingerprints to bypass two-factor authentication on high-value web services.  

\textbf{User Studies and Impact Analysis.} Redmiles et al.~\cite{RedmilesMM16} conducted extensive semi-structured interviews to understand how users prioritize security advice. Reynolds et al.~\cite{ReynoldsSBJMBE20} analyzed operational logs to quantify the impact of mandatory 2FA implementation, revealing how device remembrance, fragmented login services, and authentication timeouts contribute to user burden, and finding that this burden is comparable to other compliance tasks in large organizations. 
\section{Conclusion}

In this paper, we introduce \textbf{SE2FA}, a framework designed to assess the security of 2FA systems. Using SE2FA, we analyze the 2FA security of the top 10,000 websites from Tranco list, focusing particularly on those offering the ``Remember the Device'' functionality. Our security analysis reveals that 93 websites with this feature rely solely on cookies in their risk control processes, which is inherently insecure. Additionally, we discovered that 14 websites failed to follow secure development practices when configuring 2FA cookie attributes, leaving them vulnerable to the two specific attacks we outlined. We identified three websites with critical flaws in their 2FA implementations and assisted them in addressing these vulnerabilities.

Our study demonstrates that relying solely on cookies to enhance the usability of 2FA systems can introduce significant security risks, especially when these cookies lack adequate protective measures. We also propose mitigation strategies for affected websites and aim to raise awareness within the community about the dangers of depending exclusively on cookies to implement ``Remember the Device'' feature in 2FA systems. Finally, we have contributed our newly discovered 2FA-supported websites to the 2FA Directory to benefit the broader community.

\ifCLASSOPTIONcompsoc




%
\bibliographystyle{IEEEtran}
\bibliography{references}

\clearpage
\appendices
\section{}

\subsection{Websites Using Third-Party 2FA Solutions}
\label{sec:appendix-third-party}

Table  \ref{tab:third-party} shows the websites that use third-party accounts for login, depending on the security of the third-party 2FA systems they support. All third-party 2FA systems utilized by these websites were evaluated in our security analysis, as outlined in  Table \ref{tab2:top-500} and Table \ref{tab:top10000}. Note that these can only be accessed via third-party accounts and do not have their own account systems. 
For instance, \texttt{outlook.com} relies on a Microsoft account. Therefore, during the security assessment phase, we focused only on websites that do not depend on third-party accounts for login (G2).



\begin{table}[!ht]
    \centering
    \resizebox{0.5\textwidth}{!}{%
        \rowcolors{1}{white}{lightgray}
        \begin{tabular}{ccccc}
            \toprule
            \textbf{No.} & \textbf{Website} & \textbf{Third-Party Website} & \textbf{Attack Type} \\
            \midrule
            1 & Computers \& Technology - 1 & Shopping - 1 & A3 \\
            2 & Web-based EmailComputers \& Technology - 1 & Computers \& Technology - 1 & A3 \\
            3 & Computers \& Technology - 2 & Computers \& Technology - 1 & A3 \\
            4 & Social Networking - 1 & Social Networking - 1 & A3 \\
            5 & Search Engines \& Portals - 1 & Computers \& Technology - 1 & A3 \\
            6 & Computers \& Technology - 3 & Computers \& Technology - 1 & A3 \\
            7 & Web Phone - 1 & Computers \& Technology - 1 & A3 \\
            8 & Computers \& Technology - 4 & Computers \& Technology - 1 & A3 \\
            9 & Computers \& Technology - 5 & Computers \& Technology - 2 & A3 \\
            10 & Shopping - 1 & Shopping - 1 & A3 \\
            11 & Image SharingArts - 1 & Computers \& Technology - 2 & A3 \\
            12 & Computers \& Technology - 6 & Personal StorageComputers \& Technology - 1& A3 \\
            13 & Social Networking - 2 & Social Networking - 1 & A3 \\
            14 & Computers \& Technology - 7 & Computers \& Technology - 1 & A3 \\
            15 & Business - 1 & Computers \& Technology - 2 & A3 \\
            16 & Education - 1 & Computers \& Technology - 1 & A3 \\
            17 & Computers \& Technology - 8 & BusinessComputers \& Technology - 1 & A3 \\
            18 & Computers \& Technology - 9 & Personal StorageComputers \& Technology - 1& A3 \\
            19 & Social Networking - 3 & Social Networking - 1 & A3 \\
            20 & GamesComputers \& Technology - 1 & Computers \& Technology - 1 & A3 \\
            21 & Streaming Media \& DownloadsEntertainment - 1 & Shopping - 1 & A3 \\
            22 & Personal StorageComputers \& Technology - 1 & Computers \& Technology - 1 & A3 \\
            23 & Web-based Email - 1 & Computers \& Technology - 1 & A3 \\
            24 & Computers \& Technology - 10 & BusinessComputers \& Technology - 1 & A3 \\
            25 & Computers \& Technology - 11 & Computers \& Technology - 1 & A3 \\
            26 & Computers \& Technology - 12 & Computers \& Technology - 1 & A3 \\
            27 & Computers \& Technology - 13 & Personal StorageComputers \& Technology - 1& A3 \\
            28 & Computers \& Technology - 14 & Computers \& Technology - 1 & A3 \\
            \midrule
        \end{tabular}
    }
    \caption{Websites that rely on third-party accounts to log in and security risks associated with the third-party accounts.} 
    \label{tab:third-party}
\end{table}

\subsection{Security Evaluation}

Table \ref{tab:top10000} presents the cookies attributes and their associated attacks for the top 500 to10,00 websites that rely solely on cookies to protect their ``Remember the Device'' feature.


\begin{table*}[!hb]
    \centering
    \begin{tabular}{c|c|cccc|cc}
        \toprule
        \multirow{2}{*}{\textbf{No.}} & \multirow{2}{*}{\textbf{Website}} & \multicolumn{4}{c|}{\textbf{Cookie Attribute}} & \multirow{2}{*}{\textbf{Design Flaws (A4)}} & \multirow{2}{*}{\textbf{Attack Type}} \\
        && Amount & HTTPOnly & Secure & Expiries (days) & & \\
        \midrule
    \rowcolor{lightgray}
    24 & Pornography/Sexually Explicit - 1 & 1 & \textcolor{black}{\ding{51}} & \textcolor{black}{\ding{51}} & 400 & - & \textbf{A3} \\
    \rowcolor{lightgray}
    25 & Business - 4 & 2 & \textcolor{black}{\ding{51}} & \textcolor{black}{\ding{51}} & 365 & - & \textbf{A3} \\
    26 & Education - 1 & 1 & \textcolor{black}{\ding{51}} & \textcolor{black}{\ding{51}} & 200 & - & \textbf{A3} \\
    \rowcolor{lightgray}
    27 & Social NetworkingArts - 1 & 1 & \textcolor{black}{\ding{51}} & \textcolor{black}{\ding{51}} & 200 & - & \textbf{A3} \\
    28 & Computers \& Technology - 8 & 1 & \textcolor{black}{\ding{51}} & \textcolor{black}{\ding{51}} & 30 & - & \textbf{A3} \\
    \rowcolor{lightgray}
    29 & Social NetworkingEntertainmentBusiness - 1 & 1 & \textcolor{black}{\ding{51}} & \textcolor{black}{\ding{51}} &  \ding{116}  & - & \textbf{A3} \\
    30 & Games - 4 & 1 & \textcolor{black}{\ding{51}} & \textcolor{black}{\ding{51}} & 400 & - & \textbf{A3} \\
    \rowcolor{lightgray}
    31 & Business - 5 & 2 & \textcolor{black}{\ding{51}} & \textcolor{black}{\ding{51}} & 30 & - & \textbf{A3} \\
    32 & Computers \& Technology - 9 & 2 & \textcolor{black}{\ding{51}} & \textcolor{black}{\ding{51}} & 7 & - & \textbf{A3} \\
    \rowcolor{lightgray}
    33 & Remote Access - 1 & 1 & \textcolor{black}{\ding{51}} & \textcolor{black}{\ding{51}} & 2 & - & \textbf{A3} \\
    34 & Computers \& Technology - 10 & 1 & \textcolor{black}{\ding{51}} & \textcolor{black}{\ding{51}} & 365 & - & \textbf{A3} \\
    \rowcolor{lightgray}
    35 & Pornography/Sexually Explicit - 2 & 1 & \textcolor{black}{\ding{51}} & \textcolor{black}{\ding{51}} & 400 & - & \textbf{A3} \\
    36 & Computers \& Technology - 11 & 1 & \textcolor{black}{\ding{51}} & \textcolor{black}{\ding{51}} & 14 & - & \textbf{A3} \\
    \rowcolor{lightgray}
    37 & Computers \& Technology - 12 & 1 & \textcolor{black}{\ding{51}} & \textcolor{black}{\ding{51}} & 30 & - & \textbf{A3} \\
    38 & Computers \& Technology - 13 & 1 & \textcolor{red}{\ding{55}} & \textcolor{red}{\ding{55}} & 92 & - & \textbf{A1}, \textbf{A2}, \textbf{A3} \\
    \rowcolor{lightgray}
    39 & Computers \& Technology - 14 & 1 & \textcolor{red}{\ding{55}} & \textcolor{black}{\ding{51}} & 7 & - & \textbf{A2}, \textbf{A3} \\
    40 & Computers \& Technology - 15 & 1 & \textcolor{black}{\ding{51}} & \textcolor{red}{\ding{55}} & 30 & - & \textbf{A1}, \textbf{A3} \\
    \rowcolor{lightgray}
    41 & BusinessComputers \& Technology - 2 & 1 & \textcolor{black}{\ding{51}} & \textcolor{black}{\ding{51}} & 400 & - & \textbf{A3} \\
    42 & Games - 5 & 1 & \textcolor{black}{\ding{51}} & \textcolor{black}{\ding{51}} & 30 & - & \textbf{A3} \\
    \rowcolor{lightgray}
    43 & FinanceComputers \& Technology - 1 & 1 & \textcolor{black}{\ding{51}} & \textcolor{black}{\ding{51}} & 30 & - & \textbf{A3} \\
    44 & Education - 2 & 1 & \textcolor{black}{\ding{51}} & \textcolor{black}{\ding{51}} & 400 & - & \textbf{A3} \\
    \rowcolor{lightgray}
    45 & Computers \& Technology - 16 & 1 & \textcolor{black}{\ding{51}} & \textcolor{black}{\ding{51}} & 90 & - & \textbf{A3} \\
    46 & Computers \& Technology - 17 & 1 & \textcolor{black}{\ding{51}} & \textcolor{black}{\ding{51}} & 7 & - & \textbf{A3} \\
    \rowcolor{lightgray}
    47 & Image Sharing - 1 & 1 & \textcolor{black}{\ding{51}} & \textcolor{black}{\ding{51}} & 30 & - & \textbf{A3} \\
    48 & Computers \& TechnologyBusiness - 1 & 1 & \textcolor{black}{\ding{51}} & \textcolor{black}{\ding{51}} & 30 & - & \textbf{A3} \\
    \rowcolor{lightgray}
    49 & Computers \& Technology - 18 & 1 & \textcolor{black}{\ding{51}} & \textcolor{black}{\ding{51}} & 30 & - & \textbf{A3} \\
    50 & Information Security - 1 & 1 & \textcolor{black}{\ding{51}} & \textcolor{black}{\ding{51}} & 90 & - & \textbf{A3} \\
    \rowcolor{lightgray}
    51 & EntertainmentForums \& Newsgroups - 1 & 2 & \textcolor{black}{\ding{51}} & \textcolor{black}{\ding{51}} & 400 & - & \textbf{A3} \\
    52 & Computers \& Technology - 19 & 1 & \textcolor{black}{\ding{51}} & \textcolor{black}{\ding{51}} & 365 & - & \textbf{A3} \\
    \rowcolor{lightgray}
    53 & EntertainmentStreaming Media \& Downloads - 1 & 1 & \textcolor{black}{\ding{51}} & \textcolor{black}{\ding{51}} & 400 & - & \textbf{A3} \\
    54 & Computers \& Technology - 20 & 1 & \textcolor{black}{\ding{51}} & \textcolor{black}{\ding{51}} & 365 & - & \textbf{A3} \\
    \rowcolor{lightgray}
    55 & Computers \& Technology - 21 & 1 & \textcolor{black}{\ding{51}} & \textcolor{black}{\ding{51}} & 365 & - & \textbf{A3} \\
    56 & BusinessComputers \& Technology - 3 & 2 & \textcolor{black}{\ding{51}} & \textcolor{black}{\ding{51}} & 365 & - & \textbf{A3} \\
    \rowcolor{lightgray}
    57 & Games - 6 & 1 & \textcolor{black}{\ding{51}} & \textcolor{black}{\ding{51}} & 90 & - & \textbf{A3} \\
    58 & Computers \& TechnologyBusiness - 2 & 2 & \textcolor{black}{\ding{51}} & \textcolor{black}{\ding{51}} & 30 & - & \textbf{A3} \\
    \rowcolor{lightgray}
    59 & Real Estate - 1 & 1 & \textcolor{black}{\ding{51}} & \textcolor{red}{\ding{55}} & 400 & - & \textbf{A1}, \textbf{A3} \\
    60 & Business - 6 & 2 & \textcolor{black}{\ding{51}} & \textcolor{black}{\ding{51}} & 7 & - & \textbf{A3} \\
    \rowcolor{lightgray}
    61 & Computers \& Technology - 22 & 1 & \textcolor{black}{\ding{51}} & \textcolor{black}{\ding{51}} & 180 & - & \textbf{A3} \\
    62 & Computers \& Technology - 23 & 2 & \textcolor{black}{\ding{51}} & \textcolor{black}{\ding{51}} & 30 & - & \textbf{A3} \\
    \rowcolor{lightgray}
    63 & ShoppingComputers \& Technology - 1 & 1 & \textcolor{red}{\ding{55}} & \textcolor{black}{\ding{51}} & 400 & - & \textbf{A2}, \textbf{A3} \\
    64 & Computers \& Technology - 24 & 1 & \textcolor{black}{\ding{51}} & \textcolor{black}{\ding{51}} & 365 & - & \textbf{A3} \\
    \rowcolor{lightgray}
    65 & Computers \& Technology - 25 & 1 & \textcolor{black}{\ding{51}} & \textcolor{black}{\ding{51}} & 7 & {\color{red}\ding{108}} & \textbf{A3}, \textbf{A4} \\
    66 & Transportation - 1 & 2 & \textcolor{red}{\ding{55}} & \textcolor{red}{\ding{55}} & 30 & - & \textbf{A1}, \textbf{A2}, \textbf{A3} \\
    \rowcolor{lightgray}
    67 & Computers \& Technology - 26 & 2 & \textcolor{black}{\ding{51}} & \textcolor{black}{\ding{51}} & 30 & - & \textbf{A3} \\
    68 & Shopping - 3 & 2 & \textcolor{black}{\ding{51}} & \textcolor{black}{\ding{51}} & 7 & - & \textbf{A3} \\
    \rowcolor{lightgray}
    69 & Business - 7 & 2 & \textcolor{red}{\ding{55}} & \textcolor{red}{\ding{55}} &  \ding{116}  & - & \textbf{A1}, \textbf{A2}, \textbf{A3} \\
    70 & GamesShopping - 1 & 1 & \textcolor{black}{\ding{51}} & \textcolor{black}{\ding{51}} & 400 & - & \textbf{A3} \\
    \rowcolor{lightgray}
    71 & Government - 1 & 1 & \textcolor{black}{\ding{51}} & \textcolor{black}{\ding{51}} & 30 & - & \textbf{A3} \\
    72 & Business - 8 & 1 & \textcolor{black}{\ding{51}} & \textcolor{black}{\ding{51}} & 14 & - & \textbf{A3} \\
    \rowcolor{lightgray}
    73 & GamesForums \& Newsgroups - 2 & 1 & \textcolor{black}{\ding{51}} & \textcolor{black}{\ding{51}} & 45 & - & \textbf{A3} \\
    74 & Computers \& TechnologyBusiness - 3 & 2 & \textcolor{black}{\ding{51}} & \textcolor{black}{\ding{51}} & 30 & - & \textbf{A3} \\
    \rowcolor{lightgray}
    75 & Remote Access - 2 & 1 & \textcolor{black}{\ding{51}} & \textcolor{black}{\ding{51}} & 365 & - & \textbf{A3} \\
    76 & Real EstateComputers \& Technology - 1 & 1 & \textcolor{black}{\ding{51}} & \textcolor{black}{\ding{51}} & 7 & - & \textbf{A3} \\
    \rowcolor{lightgray}
    77 & Computers \& Technology - 27 & 1 & \textcolor{black}{\ding{51}} & \textcolor{black}{\ding{51}} & 30 & - & \textbf{A3} \\
    78 & Business - 9 & 2 & \textcolor{red}{\ding{55}} & \textcolor{black}{\ding{51}} & 30 & - & \textbf{A2}, \textbf{A3} \\
    \rowcolor{lightgray}
    79 & BusinessComputers \& Technology - 4 & 1 & \textcolor{black}{\ding{51}} & \textcolor{black}{\ding{51}} & 14 & - & \textbf{A3} \\
    80 & Shopping - 4 & 1 & \textcolor{black}{\ding{51}} & \textcolor{black}{\ding{51}} & 365 & - & \textbf{A3} \\
    \rowcolor{lightgray}
    81 & Health \& Medicine - 1 & 1 & \textcolor{black}{\ding{51}} & \textcolor{black}{\ding{51}} & 30 & - & \textbf{A3} \\
    82 & Image Sharing - 2 & 1 & \textcolor{red}{\ding{55}} & \textcolor{red}{\ding{55}} & 30 & - & \textbf{A1}, \textbf{A2}, \textbf{A3} \\
    \rowcolor{lightgray}
    83 & Computers \& Technology - 28 & 1 & \textcolor{black}{\ding{51}} & \textcolor{black}{\ding{51}} & 400 & {\color{red}\ding{108}} & \textbf{A3}, \textbf{A4} \\
    84 & Government - 2 & 1 & \textcolor{black}{\ding{51}} & \textcolor{black}{\ding{51}} & 30 & - & \textbf{A3} \\
    \rowcolor{lightgray}
    85 & Computers \& Technology - 29 & 2 & \textcolor{black}{\ding{51}} & \textcolor{black}{\ding{51}} & 45 & - & \textbf{A3} \\
    86 & Shopping - 5 & 2 & \textcolor{black}{\ding{51}} & \textcolor{black}{\ding{51}} & 365 & - & \textbf{A3} \\
    \rowcolor{lightgray}
    87 & Computers \& Technology - 30 & 1 & \textcolor{black}{\ding{51}} & \textcolor{black}{\ding{51}} &  \ding{116}  & - & \textbf{A3} \\
    88 & BusinessComputers \& Technology - 5 & 1 & \textcolor{black}{\ding{51}} & \textcolor{black}{\ding{51}} & 30 & - & \textbf{A3} \\
    \rowcolor{lightgray}
    89 & Computers \& Technology - 31 & 2 & \textcolor{black}{\ding{51}} & \textcolor{black}{\ding{51}} & 14 & - & \textbf{A3} \\
    90 & Computers \& Technology - 32 & 1 & \textcolor{black}{\ding{51}} & \textcolor{black}{\ding{51}} & 30 & - & \textbf{A3} \\
    \rowcolor{lightgray}
    91 & Computers \& Technology - 33 & 1 & \textcolor{black}{\ding{51}} & \textcolor{red}{\ding{55}} & 30 & - & \textbf{A1}, \textbf{A3} \\
    92 & BusinessComputers \& Technology - 6 & 1 & \textcolor{black}{\ding{51}} & \textcolor{black}{\ding{51}} & 365 & - & \textbf{A3} \\
    \rowcolor{lightgray}
    93 & Computers \& Technology - 34  & 1 & - & - & - & \ding{117} & \textbf{A2}, \textbf{A3} \\
    94 & Games - 7  & - & - & - & -& \ding{119} & \textbf{A4} \\
    \rowcolor{lightgray}
    95 & ShoppingEntertainment - 1  & - & - & - & - & \ding{119} & \textbf{A4} \\
    \bottomrule
    \end{tabular}
    \caption{Top 500-10,000 Websites Vulnerable  to \textbf{A1, A2, A3} and \textbf{A4} Attacks. \ding{51}:2FA Cookie attribute set to \texttt{true}, \textcolor{red}{\ding{55}}: 2FA Cookie attribute set to \texttt{false}, \ding{108}: Design flaw - cross account reuse, {\color{red}\ding{108}}: Design flaw - predictable cookie value, \ding{119}: Website 2FA is enabled but not working,  \ding{116}: 2FA cookies are session cookies. \ding{117}: Storing 2FA Cookies in \texttt{localStorage}.} 
    \label{tab:top10000}
\end{table*}

\end{document}